\documentclass[a4paper,twoside]{article}
\usepackage{amsmath,amssymb,amsfonts}
\pdfoutput=1
\usepackage{graphics}
\newtheorem{theorem}{Theorem}[section]

\begin{document}
\title{Information geometry for testing pseudorandom number generators}
\pagestyle{myheadings}
\markboth{Information geometry for testing pseudorandom number generators}{C.T.J. Dodson}

\author{ C.T.J. Dodson \\{\small\it School of Mathematics, University of Manchester,
  Manchester M13 9PL, UK}\\
  {\small\it ctdodson@manchester.ac.uk}
}

\date{}
\maketitle

\begin{abstract}
The information geometry of the 2-manifold of gamma probability density functions 
provides a framework in which pseudorandom number generators may be evaluated 
using a neighbourhood of the curve of exponential density functions. The process 
is illustrated using the pseudorandom number generator in Mathematica. This 
methodology may be useful to add to the current family of test procedures 
in real applications to finite sampling data.
\end{abstract}
\section{Introduction}
The smooth family of gamma probability density functions is given by
\begin{equation}\label{gammapdf}
    f: [0,\infty) \rightarrow [0,\infty): x \mapsto
    \frac{e^{-\frac{x \kappa }{\mu }} x^{\kappa -1} \left(\frac{\kappa }{\mu
   }\right)^{\kappa }}{\Gamma (\kappa )} \ \ \ \mu, \kappa > 0.
\end{equation}
Here $\mu$ is the mean, and the standard deviation $\sigma,$
given by $\kappa=(\frac{\mu}{\sigma})^2,$ is  proportional
to the mean. Hence the coefficient of variation $\frac{1}{\sqrt{\kappa}}$
is unity in the case that (\ref{gammapdf}) reduces to
the exponential distribution. Thus, $\kappa=1$ corresponds to an underlying Poisson
random process complementary to the exponential distribution. When $\kappa<1$ the
random variable $X$ represents spacings between events that are
more clustered than for a Poisson process and when  $\kappa>1$ the
spacings $X$ are more uniformly distributed than for Poisson.
The case when $\mu=n$ is a positive integer and $\kappa=2$ gives
the Chi-Squared distribution with $n-1$ degrees of freedom; this
is the distribution of $\frac{(n-1)s^2}{\sigma_G^2}$ for variances $s^2$
of samples of size $n$ taken from a Gaussian population with variance $\sigma_G^2.$

 The gamma distribution has a conveniently tractable information
geometry~\cite{AN,InfoGeom}, and the Riemannian metric in the
2-dimensional manifold of gamma distributions (\ref{gammapdf}) is
\begin{eqnarray}
\label{Gammametricgammakappa}
  \left[g_{ij}\right](\mu,\kappa)&=&  = \left[ \begin{array}{cc}
        \frac{\kappa}{{\mu}^2}  &    0  \\
                            0  &   \frac{d^2}{d\kappa^2}\log(\Gamma)-\frac{1}{\kappa}
\end{array} \right].
\end{eqnarray}
So the coordinates $(\mu,\kappa)$ yield an orthogonal basis
of tangent vectors, which is useful in calculations because then
the arc length function is simply
$$ds^2=\frac{\kappa}{\mu^2} \, d\gamma^2 +
        \left(
        \left(\frac{\Gamma'(\kappa)}{\Gamma(\kappa)}\right)' -
        \frac{1}{\kappa}\right)\, d\kappa^2 .$$

We note the following important uniqueness property:
\begin{theorem}[Hwang and Hu~\cite{hwang}]\label{hwangthm}
For independent positive random variables with a common probability density function $f,$
having independence of the sample mean and the sample coefficient of variation is
equivalent to $f$ being the gamma distribution.
\end{theorem}
\begin{figure}
\begin{center}
\begin{picture}(400, 200)(0, 0)
\put(0,-0){\resizebox{12 cm}{!}{\includegraphics{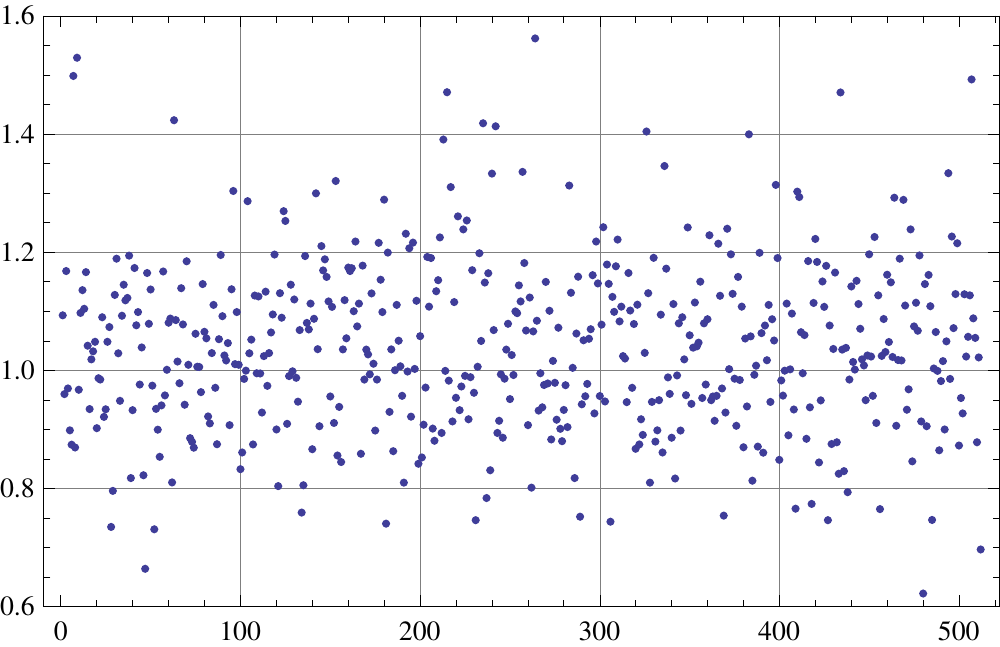}}}
\put(-5,190){{$\kappa$}}
\end{picture}
\end{center}
\caption{{\em Maximum likelihood gamma parameter $\kappa$ fitted to
separation statistics for simulations of Poisson random sequences of
length $100000$ for an element with expected parameters
$(\mu,\kappa)=(511,1).$ These simulations used the pseudorandom
number generator in Mathematica~\cite{wolfram}.} }
\label{Gamma100000}
\end{figure}
This property is one of the main reasons for the large number of applications
of gamma distributions: many near-random natural processes have standard
deviation approximately proportional to the mean~\cite{InfoGeom}. Given a set of
identically distributed, independent data values $X_1,X_2,\ldots ,X_n,$
the `maximum likelihood' or `maximum entropy' parameter values $\hat{\mu}, \hat{\kappa}$
for fitting the gamma distribution  (\ref{gammapdf})
are computed in terms of the mean and mean logarithm of the $X_i$
by maximizing the likelihood function
$$L_f(\mu,\kappa) = \prod_{i=1}^n f(X_i;\mu,\kappa).$$
By taking the logarithm and setting the gradient to zero we obtain
\begin{eqnarray}
\hat{\mu}&=&\bar{X}=\frac{1}{n}\sum^n_{i=1}X_i\\
\log\hat{\kappa} -\frac{\Gamma'(\hat{\kappa})}{\Gamma(\hat{\kappa})}
   & = & \log\bar{X} - \frac{1}{n}\sum^n_{i=1}\log X_i \nonumber \\
   & = & \log\bar{X} - \overline{\log X} . \label{maxlikgam}
\end{eqnarray}

\section{Neighbourhoods of randomness in the gamma manifold}
In a variety of contexts in cryptology for encoding, decoding or
for obscuring procedures, sequences of pseudorandom numbers are
generated. Tests for randomness of such sequences have been
studied extensively and the NIST Suite of tests~\cite{NIST} for
cryptological purposes is widely employed. Information theoretic
methods also are used, for example see Grzegorzewski and
Wieczorkowski~\cite{GW} also Ryabko and Monarev~\cite{ryabko} and
references therein for recent work. Here we can show how
pseudorandom sequences may be tested using information geometry by
using distances in the gamma manifold to compare maximum
likelihood parameters for separation statistics of sequence
elements.

{\em Mathematica}~\cite{wolfram} simulations were made of Poisson random
sequences with length $n=100000$ and spacing statistics were
computed for an element with abundance probability $p=0.00195$ in
the sequence. Figure~\ref{Gamma100000} shows maximum likelihood
gamma parameter $\kappa$ data points from such simulations. In the
data from 500 simulations the ranges of maximum likelihood gamma
distribution parameters were $419\leq\mu\leq 643$ and
$0.62\leq\kappa\leq 1.56.$

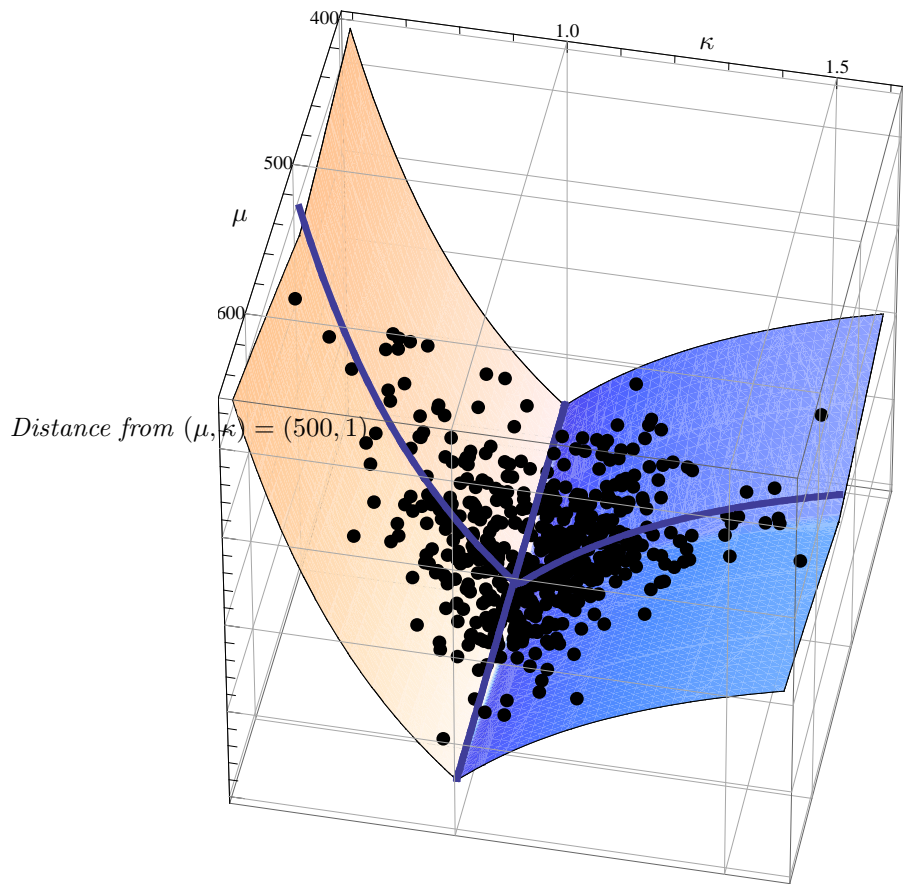
\begin{figure}
\begin{picture}(300,300)(0,0)
\put(60,0){\resizebox{9
cm}{!}{\includegraphics{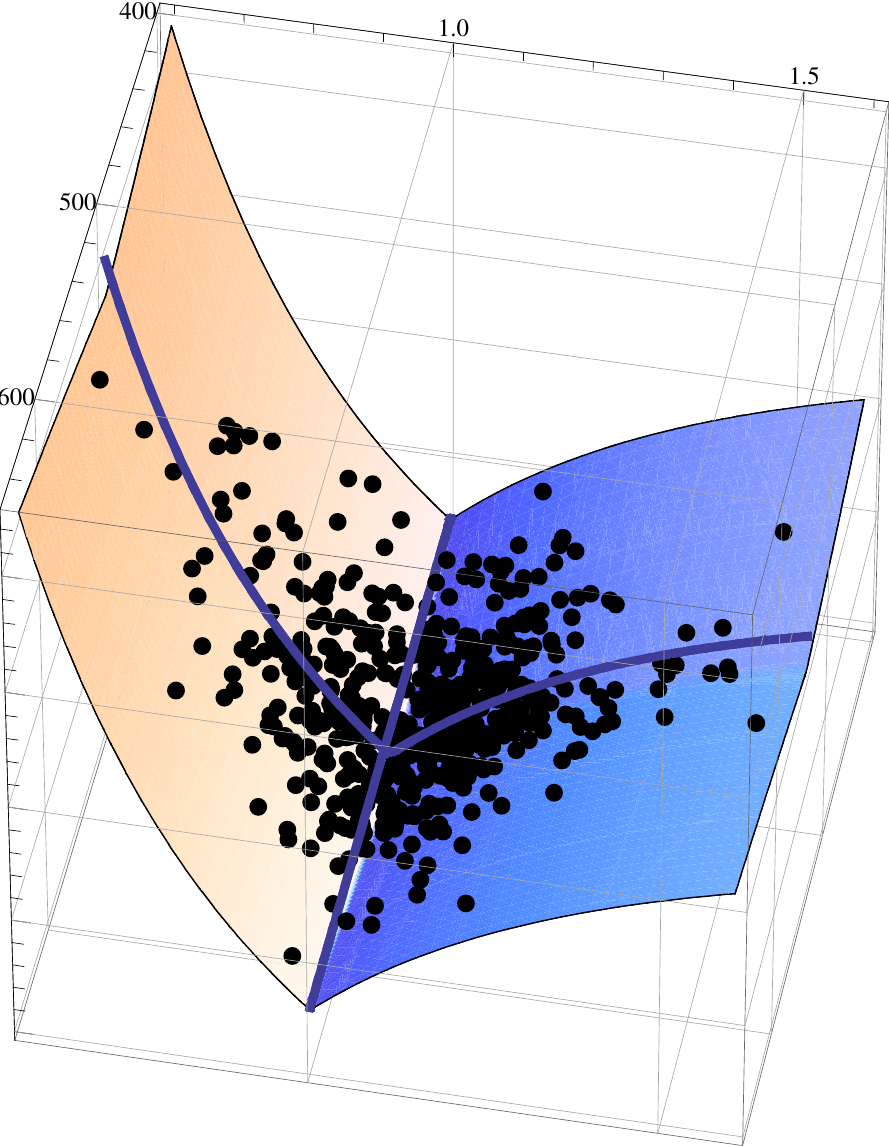}}}
\put(-18,169){\em Distance from {$(\mu,\kappa)=(500,1)$}}
\put(65,250){$\mu$}
\put(240,315){$\kappa$}
\end{picture}
\caption{{\em Distances in the space of gamma models, using a
geodesic mesh. The surface height represents upper bounds on
distances from $(\mu,\kappa)=(511,1)$ from Equation
(\ref{gddist}). Also shown are data points from simulations of
Poisson random sequences of length $100000$ for an element with expected
separation $\mu=511.$ In the limit as the sequence length tends
to infinity and the element abundance tends to zero we expect
 the gamma parameter $\kappa$ to tend to $1.$}}
\label{100000Randat}
\end{figure}

 The surface height in
Figure~\ref{100000Randat} represents upper bounds on information
geometric distances from $(\mu,\kappa)=(511,1)$ in the gamma
manifold.
 This employs the geodesic mesh function we described
in Arwini and Dodson~\cite{InfoGeom}.
\begin{equation}
Distance [(511,1),(\mu,\kappa)] \leq
\left|\frac{d^2\log\Gamma}{d\kappa^2} (\kappa) -
\frac{d^2\log\Gamma}{d\kappa^2} (1)\right| +
\left|\log{\frac{511}{\mu}}\right|. \label{gddist}
\end{equation}

Also shown in Figure~\ref{100000Randat} are
data points from the {\em Mathematica} simulations
of Poisson random sequences of length $100000$ for an element with
expected separation $\gamma=511.$

In the limit, as the sequence length tends to infinity and the
abundance of the element tends to zero, we expect
 the gamma parameter $\tau$ to tend to $1.$
However, finite sequences must be used in real applications and
then provision of a metric structure allows us, for example, to
compare real sequence generating procedures against an ideal
Poisson random model.


\end{document}